\documentclass{article}

\usepackage{graphicx,float}
\usepackage{t1enc}
\usepackage{epsfig}
\usepackage[francais,]{babel}
\usepackage{amssymb,amsbsy,amsmath,amsfonts,amssymb,amscd}

\newtheorem{remarque}{\it Remarque}
\newtheorem{exemple}{\it Exemple\/}

\begin{document}
\title{\bf R\'{e}flexions sur la question fr\'{e}quentielle en traitement du signal}

\author{Michel FLIESS \\  ~ \\ {\small{ Projet ALIEN, INRIA Futurs}} \\
\small{ \& \'Equipe MAX, LIX (CNRS, UMR 7161)}
\\  \small{\'{E}cole polytechnique, 91128 Palaiseau, France} \\
\small{ E-mail: {\tt Michel.Fliess@polytechnique.fr}}}
\date{}


\maketitle

\small{\noindent{\bf R\'{e}sum\'{e}}. On propose de nouvelles d\'{e}finitions
des
fr\'{e}quences, instantan\'{e}es ou non.  \\

\noindent{\bf Abstract}. New definitions are suggested for
frequencies, which may be instantaneous or not.

}

\section{Introduction}
Les difficult\'{e}s li\'{e}es \`{a} la transformation de Fourier en traitement
du signal sont connues depuis longtemps\footnote{Renvoyons \`a
\cite{flandrin} et \cite{max}.}. L'analyse de Fourier n'en garde pas
moins une importance capitale dans:
\begin{itemize}

\item d'innombrables applications industrielles;
\item maints d\'{e}veloppements conceptuels, comme
l'in\'{e}galit\'{e} temps-fr\'{e}quence de Hei\-senberg-Gabor ou le th\'{e}or\`{e}me
d'\'{e}chantillonnage de Shannon, apportant une armature doctrinale,
jamais remise en question.
\end{itemize}

Ces quelques pages\footnote{Elles ne doivent d'aucune fa\c{c}on \^etre
comprises comme une critique de la magnifique construction, \'{e}difi\'{e}e
par des g\'{e}ants, tels Fourier et ses continuateurs (voir
\cite{kahane}). Leur seule ambition est d'\'{e}voquer une alternative
dans un domaine tr\`{e}s d\'{e}limit\'{e}, \'{e}tranger aux p\`{e}res fondateurs.} ont
pour but de r\'{e}examiner la notion de fr\'{e}quences, dont le contenu
intuitif est ind\'{e}niable. Elles ont pour origine des publications
r\'{e}centes \cite{cr,gretsi,mexico} sur des exemples\footnote{Exemples
pour lesquels les proc\'{e}d\'{e}s issus de la lit\'{e}rature actuelle ne
donnent pas enti\`{e}re satisfaction.} de d\'{e}bruitage, de d\'{e}tection de
ruptures, de d\'{e}modulation et de compression, gr\^ace \`{a} des
techniques d'estimation de nature alg\'{e}brique\footnote{Elles sont
n\'{e}es en automatique (voir les r\'{e}f\'{e}rences de \cite{cr,gretsi,mexico})
o\`u elles sont exploit\'{e}es avec plein succ\`{e}s.}, sans lien apparent
avec l'analyse harmonique.

Une fr\'{e}quence d'un signal repr\'{e}sent\'{e} en calcul op\'{e}rationnel, au
paragraphe~\ref{operat}, par une fonction rationnelle est la partie
imaginaire d'un p\^ole. Le spectre de $\sin \omega t$, $\omega \neq
0$, $t \geq 0$, est, alors, $\pm \omega$, gr\^ace au caract\`{e}re
causal de notre approche. Celui de l'impulsion de Dirac \`{a} l'origine
est vide, contrairement \`{a} ce qui est fourni par transformation de
Fourier. Les signaux du paragaphe \ref{singular}, souvent rencontr\'{e}s
en communications, satisfont des \'{e}quations diff\'{e}rentielles
lin\'{e}aires, \`{a} coefficients polyn\^omiaux. Leurs fr\'{e}quences sont les
parties imaginaires des singularit\'{e}s des solutions des \'{e}quations
diff\'{e}rentielles op\'{e}rationnelles correspondantes. Ainsi, le spectre
du sinus cardinal $\frac{\sin \omega t}{t}$, $\omega \neq 0$, $t
\geq 0$, est encore $\pm \omega$, r\'{e}sultat qui tranche avec celui
obtenu par transformation de Fourier.

L'\'{e}cueil des \og fr\'{e}quences instantan\'{e}es \fg, deux mots {\it a
priori} antinomiques\footnote{Est-il besoin de rappeler que ce
casse-t\^ete ancien a motiv\'{e} l'introduction de superbes th\'{e}ories
nouvelles (voir \cite{flandrin,kahane,mallat,meyer}), notamment les
ondelettes ?}, est bri\`{e}vement analys\'{e} en \ref{instant}. Le
paragraphe \ref{courbure} en propose une solution \'{e}l\'{e}mentaire, bas\'{e}e
sur la courbure du signal.

In\'{e}galit\'{e} de Heisenberg-Gabor et th\'{e}or\`{e}me d'\'{e}chantillonnage de
Shannon devien\-nent, comme nous le discutons en \ref{epist}, sans
objet avec notre d\'{e}finition des fr\'{e}quences.

Ces consid\'{e}rations iconoclastes, qui heurtent des paradigmes admis
par la plupart, susciteront de vives r\'{e}sistances, quoiqu'elles aient
pour origine des techniques d'estimation ayant donn\'{e} lieu \`{a}
plusieurs demandes de brevets, d\'{e}j\`{a} d\'{e}pos\'{e}es. Nous chercherons \`{a} y
r\'{e}pondre en interrogeant les fondements pratiques du traitement du
signal et d'autres domaines scientifiques, qui reposent peu ou prou
sur les fr\'{e}quences\footnote{Le {\em Gedankenexperiment} rudimentaire
de la remarque \ref{gd} du paragraphe \ref{dirac} en est une
premi\`{e}re \'{e}bauche.}.

Nous \'{e}voquons quelques pistes nouvelles en conclusion.

\section{Calcul op\'{e}rationnel \'{e}l\'{e}mentaire}\label{operat}
\subsection{Rationalit\'{e}, p\^oles et fr\'{e}quences}\label{rat}
Soit $\mathbb{C}(s)$ le corps des fractions rationnelles en
l'ind\'{e}termin\'{e}e $s$, \`{a} coefficients complexes. Une {\em fr\'{e}quence} $f
\in \mathbb{R}$ d'un \'{e}l\'{e}ment de $\mathbb{C}(s)$ est, par d\'{e}finition,
la partie imaginaire d'un p\^ole $p = e + f \sqrt{-1}$, $e \in
\mathbb{R}$. Si $p$ est r\'{e}el, donc $f = 0$, on dit, par abus de
langage, qu'il n'y a pas de fr\'{e}quence associ\'{e}e \`{a} ce p\^ole. Le {\em
spectre} d'un \'{e}l\'{e}ment de $\mathbb{C}(s)$ est, par d\'{e}finition,
l'ensemble de ses fr\'{e}quences.
\begin{exemple}
Les deux fr\'{e}quences associ\'{e}es \`{a}
$$
\frac{\varpi}{(s - e)^2 + f^2}
$$
$f \neq 0$, o\`u $ \varpi \in \mathbb{C}[s]$ est un polyn\^ome en
l'ind\'{e}termin\'{e}e $s$, sont $\pm f$.
\end{exemple}
\begin{exemple}\label{laurent}
L'ensemble des fractions rationnelles, \`{a} p\^oles r\'{e}els, forme un
sous-anneau $R \subset \mathbb{C}(s)$, qui contient l'anneau
$\mathbb{C}[s, s^{-1}]$ des polyn\^omes de Laurent, c'est-\`{a}-dire les
sommes finies $\sum_{\alpha \in \mathbb{Z}} c_\alpha s^\alpha$,
$c_\alpha \in \mathbb{C}$. On dira que le spectre de tout \'{e}l\'{e}ment de
$R$ est vide.
\end{exemple}

\subsection{Polyn\^omes exponentiels}\label{polexp}
Appelons {\em polyn\^ome exponentiel} toute somme finie
\begin{equation}
\label{exp} \sum_{\iota = 1, \dots, N} P_\iota(t) e^{a_\iota t}
\end{equation}
$a_\iota \in \mathbb{C}$, o\`u $P_\iota \in \mathbb{C}[t]$ est un
polyn\^ome en la variable r\'{e}elle $t$. Pour $t \geq 0$, le calcul
op\'{e}rationnel\footnote{Le calcul op\'{e}rationnel est traditionnellement
abord\'{e} avec la transformation de Laplace ({\it cf.} \cite{max}).
L'approche alg\'{e}brique de Mikusi\'nski, ind\'{e}pendante de cette
transformation, nous semble pr\'{e}senter de gros avantages (voir
\cite{miku1,miku2,yosida}).} \'{e}tablit une bijection entre polyn\^omes
exponentiels et fractions rationnelles de $\mathbb{C}(s)$,
strictement propres, c'est-\`{a}-dire avec degr\'{e}s des num\'{e}rateurs
strictement inf\'{e}rieurs \`{a} ceux des d\'{e}nominateurs. Il en d\'{e}coule que
le spectre de (\ref{exp}) est form\'{e} des parties imaginaires des
coefficients $a_1, \dots, a_N$, car ceux-ci sont les p\^oles de la
fraction rationnelle correspondante.
\begin{exemple}
Le spectre de la fonction polyn\^omiale $P(t)$, $P \in
\mathbb{C}[t]$, $t \geq 0$, est vide. Celui de la fonction $P(t)
\sin (\omega t + \phi)$, $P \neq 0$, $\omega, \phi \in \mathbb{R}$,
$\omega \neq 0$, $t \geq 0$, est form\'{e} de $\pm \omega$.
\end{exemple}

\begin{remarque}
Ces r\'{e}sultats diff\`{e}rent de ceux obtenus avec la transformation de
Fourier, o\`u, en particulier, $\sin (\omega t + \phi)$, $t \geq 0$,
poss\`{e}de toutes les fr\'{e}quences. Que l'on nous permette d'y voir un
atout de notre point de vue, d\^u au caract\`{e}re {\em causal} du
calcul op\'{e}rationnel.
\end{remarque}

\subsection{Impulsion de Dirac}\label{dirac}
L'impulsion, ou mesure, de Dirac \`{a} l'origine, $\delta$, correspond \`{a}
$1$ dans $\mathbb{C}(s)$. Son spectre est donc vide.
\begin{remarque}
Rappelons que la transform\'{e}e de Fourier de $\delta$ vaut \'{e}galement
$1$. Il en d\'{e}coule, dans ce langage, que le spectre de $\delta$
poss\`{e}de toutes les fr\'{e}quences, qui \'{e}videmment se {\em d\'{e}truisent en
se compensant} hors de l'origine.
\end{remarque}
\begin{remarque}\label{gd}
Prenons, avec les ing\'{e}nieurs et les physiciens, un filtre {\em
r\'{e}alisable}, c'est-\`{a}-dire \`{a} {\em fonction de transfert}
$\mathfrak{F}$, {\em rationnelle et propre}, approximation d'un
filtre \`{a} {\em bande \'{e}troite} autour des fr\'{e}quences $\pm \omega_0$,
$\omega_0 \neq 0$ ({\it cf.} \cite{max}). Sa {\em r\'{e}ponse
impulsionnelle} $\mathfrak{R}$, c'est-\`{a}-dire sa r\'{e}ponse \`{a}
l'impulsion de Dirac\footnote{Bien entendu, on utilise en pratique
une approximation.}, ne signifie pas, d'apr\`{e}s nous, que $\delta$
contient les fr\'{e}quences $\pm \omega_0$, mais traduit un fait
alg\'{e}brique trivial, le produit de $\mathfrak{F}$ par $1$,
c'est-\`{a}-dire la convolution de $\mathfrak{R}$ avec l'\'{e}l\'{e}ment neutre.
\end{remarque}

\section{Fr\'{e}quences et singularit\'{e}s}\label{singular}
\subsection{D\'{e}rivation alg\'{e}brique}
Munissons $\mathbb{C}(s)$ d'une structure de {\em corps
diff\'{e}rentiel} (voir, par exemple, \cite{loir,put}) avec la
d\'{e}rivation, dite {\em alg\'{e}brique} \cite{miku1,miku2},
$\frac{d}{ds}$. On sait \cite{miku1,miku2,yosida} que cette
d\'{e}rivation correspond \`{a} la multiplication par $-t$ des polyn\^omes
exponentiels. L'anneau non commutatif des op\'{e}rateurs diff\'{e}rentiels
lin\'{e}aires $\mathbb{C}(s)[\frac{d}{ds}]$, \`{a} coefficients rationnels,
est principal \`{a} gauche et \`{a} droite et contient l'alg\`{e}bre de Weyl
$\mathbb{C}[s, \frac{d}{ds}]$ \cite{mc}.

Un signal\footnote{Comparer avec \cite{cr,gretsi,mexico}.} $x
\not\in \mathbb{C}(s)$ examin\'{e} ci-dessous est solution d'une
\'{e}quation diff\'{e}rentielle lin\'{e}aires non n\'{e}cessairement homog\`{e}ne : il
existe $L \in \mathbb{C}(s)[\frac{d}{ds}]$, $L \not\in
\mathbb{C}(s)$, $\varpi \in \mathbb{C}(s)$ tels que $L x = \varpi$.
Les techniques traditionnelles d'alg\`{e}bre diff\'{e}rentielle
\cite{loir,put} permettent de consid\'{e}rer $x$ comme appartenant \`{a} une
{\em extension de Picard-Vessiot} de $\mathbb{C} (s)$.

Les singularit\'{e}s des solutions d'\'{e}quations diff\'{e}rentielles lin\'{e}aires
(voir, par exemple, \cite{put}) permettent de g\'{e}n\'{e}raliser \ref{rat}:
une {\em fr\'{e}quence} de $x$ est la partie imaginaire $f \in
\mathbb{R}$ d'une singularit\'{e} $e + f \sqrt{- 1}$, $e \in
\mathbb{R}$, de $x$. Le {\em spectre} de $x$ est, ici encore,
l'ensemble des fr\'{e}quences.

\begin{remarque}
Cette approche du calcul op\'{e}rationnel, qui semble nouvelle, \'{e}vite
l'usage des tables de transformations, comme \cite{ditkin}.
\end{remarque}

\subsection{Quelques signaux courants}
\subsubsection{Sinus cardinal}\label{sinc} Le sinus cardinal $\sigma = \frac{\sin
\omega t}{t}$, $\omega \in \mathbb{R}$, $\omega \gneqq 0$, $t \geq
0$, satisfait\footnote{Ici, comme en \ref{cos} et \ref{infini}, on
d\'{e}signera avec les m\^emes notations le signal, fonction de $t$, et
son image en calcul op\'{e}rationnel.}
$$
\frac{d \sigma}{ds} + \frac{\omega}{s^2 + \omega^2} = 0
$$
Il est clair que $\sigma$ poss\`{e}de deux singularit\'{e}s {\em
logarithmiques} \cite{hurwitz} en $\pm \omega \sqrt{- 1}$. Le
spectre de $\sigma$ est form\'{e} de $\pm \omega$.

\begin{remarque}
Rappelons que la transform\'{e}e de Fourier de $\frac{\sin \omega
t}{t}$, $t \in \mathbb{R}$, est
\begin{equation}\label{cardinal}
\chi_\omega (\xi) = \left\{ \begin{array}{l} \omega \quad \mbox{\rm
si} ~ - \omega < \xi < \omega \\ 0 \quad \mbox{\rm si} ~  \xi < -
\omega ~ \mbox{\rm ou} ~ \xi > \omega
\end{array} \right.
\end{equation}
\end{remarque}

\subsubsection{Cosinus sur\'{e}lev\'{e}}\label{cos} Le cosinus sur\'{e}lev\'{e} $\varrho =
\frac{\cos \omega t}{t^2 + 1}$, $\omega \in \mathbb{R}$, $\omega
\gneqq 0$, $t \geq 0$, satisfait
$$
\left(\frac{d^2}{ds^2} + 1 \right) \varrho = \frac{s}{s^2 +
\omega^2}
$$
On v\'{e}rifie ais\'{e}ment que le spectre est form\'{e} de $\pm \omega$.

\subsubsection{Retards et avances}
L'exponentielle op\'{e}rationnelle $\varrho = e^{-Ls}$, $L \in
\mathbb{R}$, d\'{e}signe l'op\'{e}rateur de retard si $L >0$, d'avance si $L
< 0$. Elle satisfait l'\'{e}quation
$$
\left( \frac{d}{ds} + L \right) \varrho = 0
$$
sans singularit\'{e}s. Ce spectre vide est en accord avec \ref{dirac}.

\subsubsection{Singularit\'{e}s \`{a} l'infini}\label{infini} Le signal
complexe
$$\varepsilon = \exp \left[ (a t^2 + b
t  + c ) \sqrt{-1}\right]$$ $a, b, c \in \mathbb{R}$, $a \neq 0$,
satisfait l'\'{e}quation
$$
\left[ s + \left( 2a \frac{d}{ds} - b \right) \sqrt{-1} \right]
\varepsilon = \exp ( c \sqrt{-1} )
$$
sans singularit\'{e} finie. Le spectre, au sens entendu ici, est donc
vide. Il existe par contre une singularit\'{e} infinie \cite{put}, qui
traduit les oscillations de plus en plus {\it rapides} de
$\varepsilon$ lorsque $t \to + \infty$. Il serait peut-\^etre
int\'{e}ressant d'enrichir la notion de spectre afin d'y inclure ce type
de comportement\footnote{Les travaux de Poincar\'{e}, Birkhoff,
Turrittin, et d'autres (voir, en \cite{put}, le th\'{e}or\`{e}me 3.1 de  et
ses divers prolongements) devraient y jouer un r\^ole majeur. Ces
m\^emes r\'{e}sultats, appliqu\'{e}s dans le domaine temporel, aux \'{e}quations
\`{a} coefficients analytiques devraient permettre de d\'{e}finir leurs
fr\'{e}quences. Quant aux signaux non lin\'{e}aires, c'est-\`{a}-dire
satisfaisant des \'{e}quations diff\'{e}rentielles non lin\'{e}aires, sugg\'{e}rons
que leurs spectres pourraient se d\'{e}finir gr\^ace aux \'{e}quations
lin\'{e}aires variationnelles.}.

\section{Fr\'{e}quences instantan\'{e}es}
\subsection{Dur\'{e}e limit\'{e}e}\label{instant}
La d\'{e}finition suivante comble certains manques de l'approche
traditionnelle: le spectre d'un signal $x$, co\"{\i}ncidant, sur un
intervalle de temps arbitraire $[t_0, t_0 + h[$, $h > 0$, avec l'un
de ceux examin\'{e}s aux paragraphes \ref{operat} et \ref{singular},
soit $\xi$, est celui de $\xi$. Elle n'est cependant pas
satisfaisante:

Supposons, pour simplifier, $\xi = \sin \omega t$, $\omega \neq 0$.
Il para\^{\i}t impossible, si $h$ est {\it petit}, de distinguer $x$ de
son d\'{e}veloppement de Taylor tronqu\'{e} \`{a} un ordre {\em suffisamment
\'{e}lev\'{e}}, dont, d'apr\`{e}s \ref{polexp}, le spectre est vide.

On obtiendrait ainsi deux r\'{e}ponses contradictoires.

\subsection{Cercle osculateur}\label{courbure}
Supposons le signal $t \mapsto x(t)$ localement $C^2$. En {\it
confondant} autour du point $(t, x(t))$ son graphe de courbure
$$\frac{ \ddot{x}(t)}{\left( 1 +
(\dot{x}(t))^2 \right)^{\frac{3}{2}}}$$ avec le cercle osculateur,
on obtient la {\em fr\'{e}quence instantan\'{e}e} en $t$:
\begin{equation*}
\label{freq} \Phi (t) = \frac{ \ddot{x}(t) }{ \sqrt{ 1 +
(\dot{x}(t))^2 }}
\end{equation*}
Il est imm\'{e}diat de v\'{e}rifier que cette d\'{e}finition, tr\`{e}s large, ne
fournit pas les m\^emes r\'{e}sultats que celle de Ville ({\it cf.}
\cite{flandrin,max}).
\begin{exemple}
La fr\'{e}quence instantan\'{e}e d'un signal constant par morceaux est nulle
preque partout.
\end{exemple}
\begin{exemple}
Pour $x(t) = A \sin \omega t$, $A, \omega \in \mathbb{R}$, $\omega
\neq 0$, il vient\footnote{Si l'on sait {\it a priori} avoir \`{a} faire
\`{a} un signal sinuso\"{\i}dal $A \sin (\omega t + \varphi)$, on peut
estimer $A, \omega, \varphi \in \mathbb{R}$ sur une fen\^etre de
temps tr\`{e}s courte, y compris dans un environnement bruit\'{e} (voir
\cite{mexico}).}
$$
\Phi (t) = \frac{ \omega^2 A \sin \omega t }{\sqrt{ 1 + \omega^2 A^2
\cos^2 \omega t }}
$$
\end{exemple}

\section{Cons\'{e}quences m\'{e}thodologiques}\label{epist}
\subsection{In\'{e}galit\'{e} de Heisenberg-Gabor}
Illustrons le cadre habituel de l'in\'{e}galit\'{e} de Heisenberg-Gabor par
les deux exemples suivants:
\begin{enumerate}
\item Le support de l'impulsion de Dirac $\delta$ est ponctuel,
alors que, sa transform\'{e}e de Fourier \'{e}tant \'{e}gale \`{a} $1$, son spectre
est uniform\'{e}ment r\'{e}parti sur $\mathbb{R}$.
\item On sait que la limite, pour $\omega \to \pm \infty$, du sinus cardinal $\frac{\sin
\omega t}{\omega t}$, $t \in \mathbb{R}$, est $\delta$. La formule
(\ref{cardinal}) indique un \'{e}talement de plus en plus grand du
spectre.
\end{enumerate}
Les calculs de \ref{dirac} et \ref{sinc} montrent que cette
in\'{e}galit\'{e} temps-fr\'{e}quence perd tout sens ici.

\subsection{\'Echantillonnage}
Il est impossible de trouver un analogue au th\'{e}or\`{e}me
d'\'{e}chantillonnage de Shannon, qui, rappelons-le, est acausal, car il
repose sur la transformation de Fourier. Pour un signal transitoire
arbitraire\footnote{On aura compris qu'on ne cherche pas ici \`{a}
d\'{e}finir le spectre d'un tel signal.}, notre d\'{e}marche \cite{cr}
conduit \`{a} des proc\'{e}d\'{e}s voisins de l'analyse num\'{e}rique classique: on
estime les d\'{e}riv\'{e}es du signal\footnote{Insistons sur le fait que
cette estimation fonctionne avec des signaux bruit\'{e}s.} jusqu'\`{a} un
certain ordre et on utilise les techniques usuelles
d'interpolation\footnote{Unser \cite{unser1,unser2} pr\^one aussi la
consid\'{e}ration des signaux physiques, c'est-\`{a}-dire continus, avant
tout \'{e}chantillonnage, pour lequel il propose certains types de
splines.}.

\section{Conclusion}
La citation suivante, due \`{a} de Broglie \cite{broglie}, est bien
connue:

\noindent{\it \og La consid\'{e}ration exclusive des ondes
monochromatiques conduit \`{a} une autre conception qui me para\^{\i}t
erron\'{e}e. Si l'on consid\`{e}re une grandeur qui peut \^etre repr\'{e}sent\'{e}e,
\`{a} la mani\`{e}re de Fourier, par une superposition de composantes
monochromatiques, c'est la superposition qui a un sens physique et
non les composantes de Fourier consid\'{e}r\'{e}es isol\'{e}ment. \fg}

\noindent Elle sugg\`{e}re que certains des paradoxes de
l'interpr\'{e}tation, dite \og orthodoxe \fg, de la m\'{e}canique quantique
pourraient \^etre dues \`{a} des manipulations hasardeuses des
fr\'{e}quences.

In\'{e}galit\'{e} de Heisenberg-Gabor et th\'{e}or\`{e}me d'\'{e}chantillonnage de
Shannon expriment les limitations inh\'{e}rentes \`a toute analyse d'un
signal. Nous reviendrons sur ce point incontournable en r\'{e}examinant
la notion de {\em bruit}\footnote{Rappelons que nos techniques
d'estimation \cite{cr,gretsi,mexico} sont ind\'{e}pendantes de toute
consid\'{e}ration statistique.}. L'analyse non standard ({\em cf.}
\cite{robinson}) devrait nous y aider.



\end{document}